\documentclass[conference, a4paper]{IEEEtran}
\IEEEoverridecommandlockouts

\ifCLASSINFOpdf
  \usepackage[pdftex]{graphicx}
  \DeclareGraphicsExtensions{.pdf,.jpeg,.png}
\else
\fi

\usepackage{cite}
\usepackage{amsmath,amssymb,amsfonts}
\usepackage{algorithmic}
\usepackage{graphicx}
\usepackage{textcomp}
\usepackage[left=1.4cm, right=1.4cm, top=1.7cm]{geometry}
\usepackage{caption}
\usepackage{anyfontsize}
\usepackage{xcolor}
\usepackage{subcaption}
\usepackage[pagebackref=true]{hyperref}
\usepackage{multirow}
\usepackage{hyperref}
\usepackage{url}
\usepackage{flushend}
\makeatletter
\newcommand*\titleheader[1]{\gdef\@titleheader{#1}}
\AtBeginDocument{%
\let\st@red@title\@title
\def\@title{%
\bgroup\normalfont\normalsize\centering\@titleheader\par\egroup
\vskip0.2em\st@red@title}
}
\makeatother

\makeatletter
\renewcommand{\fnum@figure}{Figure \thefigure}
\makeatother

\title{ {Decentralized Autonomous Traffic Management through Corridor Networks} \\
\thanks{This work was supported in part by NASA grant \#80NSSC23M0220, but this article solely reflects the opinions and conclusions of its authors and not any NASA entity. This research was also sponsored by the Department of the Air Force Artificial Intelligence Accelerator and was accomplished under Cooperative Agreement Number FA8750-19-2-1000. The views and conclusions contained in this document are those of the authors and should not be interpreted as representing the official policies, either expressed or implied, of the Department of the Air Force or the U.S. Government. The U.S. Government is authorized to reproduce and distribute reprints for Government purposes notwithstanding any copyright notation herein.}
\vspace{0.5cm}
}

\titleheader{Second US-Europe Air Transportation Research and Development Symposium (ATRDS2026)}

\author{\IEEEauthorblockN{Jasmine Jerry Aloor}
\IEEEauthorblockA{Department of Aeronautics and Astronautics \\
Massachusetts Institute of Technology \\
Cambridge, MA, USA \\
jjaloor@mit.edu}
\and
\IEEEauthorblockN{Aadarsh Govada }
\IEEEauthorblockA{
University of Maryland \\
College Park, MD, USA \\
agovada@terpmail.umd.edu} 
\and 
\IEEEauthorblockN{Hamsa Balakrishnan}
\IEEEauthorblockA{Department of Aeronautics and Astronautics \\
Massachusetts Institute of Technology \\
Cambridge, MA, USA \\
hamsa@mit.edu}  
}

\renewcommand\IEEEkeywordsname{Keywords}
\IEEEaftertitletext{\vspace{-1\baselineskip}}

\begin{document}

\maketitle

\noindent \begin{abstract}
As autonomous aircraft are introduced at scale and traffic density increases, centralized management becomes insufficient to coordinate the large numbers of crewed and uncrewed aircraft. Dedicated Advanced Air Mobility (AAM) corridors have therefore been proposed for organizing high-density autonomous traffic flows. The desire to scalably provide autonomous aircraft flexibility in trajectory planning motivates the development of decentralized approaches to traffic management in AAM corridors.

In this work, we extend a multi-agent reinforcement learning (MARL) approach to address the challenge of decentralized traffic flow management in air corridor networks. We test policies trained in a single-corridor setting on increasingly complex multi-corridor networks with combinations of merges and splits in a zero-shot manner. Experimental results demonstrate that learned behaviors transfer well to scenarios with varying traffic density, network geometry, and heterogeneous vehicle performance, without needing centralized coordination or model retraining. We evaluate system-level performance in terms of conformance to corridor boundaries, completion rates, average speeds, distance traveled, and maintenance of inter-aircraft separation. We find that although our policies require only locally coordinated entry, traversal, and exit behaviors, they collectively produce desirable traffic flows through the corridor network.
\end{abstract}

\vspace{0.3cm}

\begin{IEEEkeywords}
advanced air mobility; corridors; decentralized traffic management; multi-agent reinforcement learning; traffic flow management
\end{IEEEkeywords}

\section{Introduction}

Advanced Air Mobility (AAM) is expected to introduce large numbers of autonomous aircraft operating in dense airspace. The advent of these new types of aircraft will be accompanied by fundamental challenges, including vastly increased traffic demand, communication bottlenecks, single points of failure, and limited responsiveness to local traffic conditions \cite{US_DOT2025_AAM_National_strategy, US_DOT2025_AAM_Comprehensive_Plan, schuchardt_uam_assessment_2026}. Safe and efficient operations at scale, therefore, call for decentralized approaches that distribute decision-making across autonomous agents while maintaining inter-aircraft separation.

\begin{figure}[!b]
    \centering
    \includegraphics[width=0.55\columnwidth]{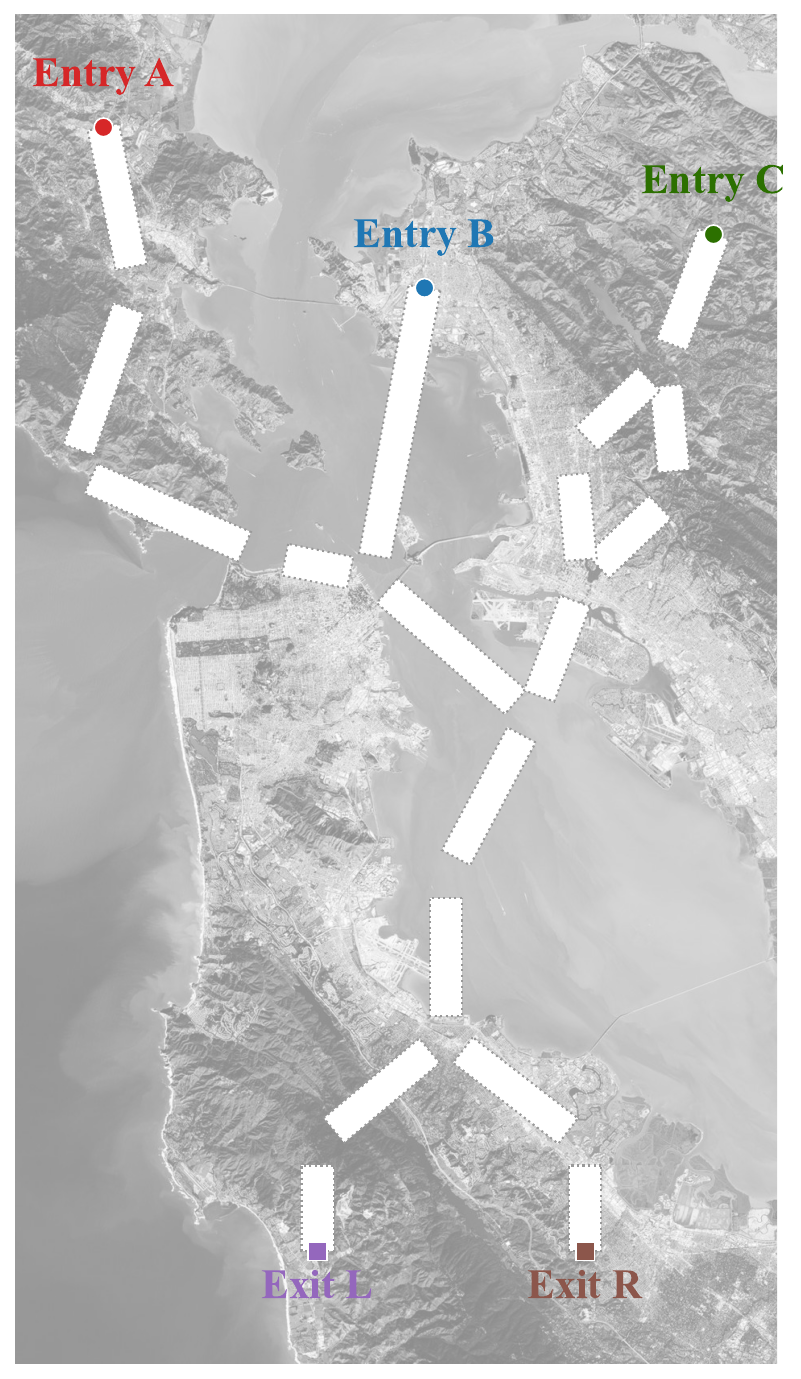}
    \caption{Schematic of a corridor network, showing the layout of routes, splits, and merges. There are three entry points to the network at the top of the image, and two exit points at the bottom. This figure is for illustrative purposes only.}
    \label{fig:SFBaycorriday}
\end{figure}
Recent works have suggested dedicated AAM corridors as a promising concept for structuring traffic flows and reducing coordination complexity in high-density environments \cite{aloor_decentralized_2026, homola_corridors_2012}. Corridors organize aircraft into predefined routes in order to enable predictable motion while supporting decentralized coordination. Within these structured environments, aircraft must maintain separation, conform to corridor boundaries, and minimize congestion-induced delays. Consequently, corridor-based operations have become an important setting for the development of decentralized traffic flow coordination strategies. Fig. \ref{fig:SFBaycorriday} illustrates the schematic of a possible AAM corridor network with three entry points, two exits, 18 corridor segments, and eight possible routes.

Multi-agent reinforcement learning (MARL) has demonstrated strong potential for learning cooperative behaviors in complex multi-vehicle systems \cite{nayak_scalable_2023, yu_transformer_2025}, enabling agents to discover coordination strategies through environmental interaction rather than relying on hand-crafted rules. While prior work has demonstrated the feasibility of decentralized learning-based coordination, many studies focus on low traffic densities or simplified scenario structures \cite {brittain_attention_2023,brittain_heterogeneous_2022,aloor_decentralized_2026}. However, the development of coordination policies that can handle complex corridor topologies, scale to larger numbers of agents and more sustained traffic flows, and accommodate heterogeneous agent behavior has remained an open challenge.

In this work, we extend and evaluate our prior decentralized MARL framework for AAM corridor operations \cite{aloor_decentralized_2026}, by introducing three key methodological advances: (a) a rotation-invariant policy representation, (b) curriculum-based training to enforce robust corridor conformance, and (c) evaluation of zero-shot transfer to complex multi-corridor networks. Our policies do not require that the aircraft have information on the entire network as it begins its trajectory; at any time, it only has information on its next corridor entry and exit point, along with the orientation of the associated corridor. A rotation-invariant coordination policy trained in a single-corridor environment is deployed zero-shot in multi-corridor settings---including merging, branching, and combined layouts---without any retraining or centralized scheduling. We evaluate performance across increasing traffic densities and mixed fleets of vehicles with heterogeneous performance envelopes, analyzing system-level effects such as throughput efficiency and corridor conformance. Our results show that locally learned behaviors generalize well across corridor topologies and scale with traffic demand, suggesting that structured airspace design and decentralized policy learning are complementary approaches to scalable AAM traffic management. It is important to note that the developed methods fall into the realm of \emph{traffic flow management}: the objective is for agents to learn good strategic behavior such as conformance to structure and congestion management. While there would still be a need for tactical interventions to maintain separation (e.g., through sense-and avoid, similar to the role played by air traffic control today), we wish for these interventions to be relatively infrequent, even in congested scenarios. Our evaluations in Sec. \ref{sec:results} show that indeed, such tactical interventions are needed less than 5\% of the time, even in congested complex corridor scenarios.

The remainder of this paper is organized as follows: Section \ref{sec:related} positions our paper within related research areas. Section \ref{sec:practical} positions the work within the broader AAM traffic management context and describes the operational and technological assumptions that underpin it. Section \ref{sec:method} describes our aircraft models, environment, and MARL training methodology. Section \ref{sec:experiment} presents the evaluation setup using the trained models. Section \ref{sec:results} discusses the key results, Section \ref{sec:limit} discusses some limitations of our approach and directions for future work, and Section \ref{sec:conc} concludes the paper.
Materials and source code corresponding to this paper are available online\footnote{Code:
\url{https://github.com/Jaroan/Decentralized-Corridor-MARL}.
Project page: \url{https://jaroan.github.io/jasminejerrya/AAM_Corridor_MARL.html}.}.

\section{Related Work}
\label{sec:related}
\subsection{Centralized traffic management in AAM}
Motivated by conventional air traffic flow management approaches, a considerable portion of prior research in AAM traffic management has focused on centralized or system-level coordination strategies aimed at ensuring efficiency and safety at the network scale. National AAM planning efforts emphasize structured operational concepts and coordinated traffic flow management to support the integration of large numbers of autonomous aircraft into existing airspace systems \cite{US_DOT2025_AAM_National_strategy,US_DOT2025_AAM_Comprehensive_Plan,FAA_UAM_CONOPS_2023, schuchardt_uam_assessment_2026,
kopardekar_uas_operations_2014}. Recognizing the benefits of decreased centralization, frameworks such as U-Space \cite{SESARJointUndertaking} and UTM \cite{FAA_UAM_CONOPS_2023} adopt federated constructs that distribute responsibilities across service providers. At the algorithmic level, centralized scheduling, hybrid control architectures, and distributed traffic flow management frameworks leverage global system information to regulate congestion and maintain orderly traffic patterns \cite{tomlin_hybrid_1996,balakrishnan_distributed_2017} and aggregate modeling approaches have been proposed to capture traffic dynamics in structured urban airspace based on system-level occupancy and throughput \cite{chour_mfd_2024}. 
Hierarchical decentralized approaches grounded in formal methods have also been proposed, in which ground infrastructure retains local control authority and runtime shielding provides formal safety guarantees~\cite{bharadwaj_decentralized_2021}.
These methods provide global insights into traffic flows but rely on system-wide state estimation, which limits responsiveness to fast-evolving local interactions in dense corridor environments.

\subsection{Corridor-based operations}

Dedicated corridor concepts have emerged as a prominent approach for organizing AAM traffic within dense urban airspace. Operational planning documents and recent research emphasize the use of structured routes to improve predictability, reduce conflict complexity, and enable scalable integration of autonomous aircraft into existing airspace systems\cite{US_DOT2025_AAM_National_strategy,US_DOT2025_AAM_Comprehensive_Plan, SESARJointUndertaking, verma_corridor_design_2022, homola_corridors_2012}. By constraining aircraft motion to predefined pathways, corridors introduce an intermediate layer between fully centralized air traffic control and fully unstructured free-flight operations.

Prior studies have investigated corridor geometry, compliance strategies, and planning methodologies to support structured traffic flows. Design frameworks explore how lane structure and network-level planning can influence traffic efficiency and safety outcomes\cite{tony_corridrone_2021,he_corridor_planning_2025}. These structured environments introduce interaction patterns such as merging flows, sequential corridor transitions, and branching layouts that differ significantly from traditional airspace scenarios. While corridor-based operations can simplify global planning, they also shift coordination challenges toward local interactions among nearby aircraft, motivating decentralized and distributed approaches to managing structured AAM traffic.


\subsection{Decentralized coordination}

Decentralized coordination has long been explored as an alternative to centralized air traffic management, particularly in environments where local interactions dominate system behavior. Prior work has investigated distributed control strategies and self-organizing mechanisms that enable autonomous agents to coordinate using locally available information. Adaptive traffic-following approaches demonstrate how distributed multi-vehicle systems can dynamically balance order and efficiency as traffic density changes, highlighting the potential benefits of self-organizing behaviors in scalable airspace operations \cite{jain_traffic_following_2025}. Geometric and rule-based self-separation methods further illustrate how aircraft can maintain safe interactions without continuous centralized oversight \cite{liu_self_separation_2025}, while structured policies such as merge-assist mechanisms have been proposed to manage conflicts in constrained urban airspace\cite{doole_merge_assist_2022}.

Beyond aviation-specific applications, related research in convoy and platoon coordination shows how distributed agents can achieve collective behavior through local interaction rules rather than global control \cite{henke_convoy_2006,heinovski_platoon_2018,johansson_platoon_2022,tsugawa_truck_platoon_2016}. These approaches provide interpretable and operationally grounded solutions, but their reliance on predefined coordination logic can limit adaptability across diverse traffic densities, heterogeneous agent behaviors, and complex corridor topologies. 
Unlike geometric or rule-based self-separation approaches, which rely on explicitly designed conflict resolution logic, our method learns coordination behaviors directly from interaction data.

\subsection{Learning-based decentralized coordination in AAM corridors}

Prior studies have investigated reinforcement learning for conflict resolution and cooperative multi-agent navigation, demonstrating the potential for decentralized decision-making in aviation and structured mobility settings \cite{brittain_attention_2023,brittain_heterogeneous_2022,yu_transformer_2025, aloor_decentralized_2026, AloorJ-RSS-25, ahmad2026hmarl}, including
attention-based architectures that handle variable numbers of neighbors and
policies designed for heterogeneous agent objectives.


While these studies illustrate the feasibility of learning-based decentralized coordination, evaluations are often conducted under limited traffic densities or simplified scenario structures. As a result, the ability of decentralized MARL policies to scale across large agent populations, heterogeneous behaviors, and complex corridor topologies remains an open question, motivating the focus of the present work.

\section{Operational and Technological Assumptions}
\label{sec:practical}

Before describing the methodology, it is important to describe assumptions made regarding the operating environment and technologies. These assumptions also influence the model and formulation presented in Section~\ref{sec:method}.

\subsection{Operational context}
AAM traffic management spans a spectrum from strategic scheduling and demand-capacity balancing, to tactical conflict resolution acting on imminent loss-of-separation events. The coordination approach studied in this work sits between these layers: agents act continuously on local observations and issue guidance-level commands that shape how aircraft traverse and merge within corridors, rather than scheduling departures upstream or  resolving last-second conflicts. Within an operational system, the learned policy is best understood as a flow-shaping layer in advance of a tactical safety mechanism that intervenes only when a loss of separation becomes imminent. The tactical safety mechanism, which is not the focus of this paper, would provide a last-resort guarantee that is independent of the learned behavior. In other recent work, we have developed such a layer for the corridor merging setting, enforcing collision avoidance using a runtime safety filter based on Hamilton-Jacobi reachability value functions while minimally modifying the reference guidance \cite{low2026timetoreach}. Pairing the policy studied in this paper with a tactical filter---whether reachability-based as in~\cite{low2026timetoreach} or barrier-function-based as in~\cite{ahmad2026hmarl}---offers a concrete path to combining the scalability of learned decentralized coordination with formal separation guarantees.

\subsection{Information sharing and level of decentralization}
In this work, we adopt a strongly decentralized architecture. During the time of execution, no central controller computes or distributes trajectories, and agents do not exchange intent, negotiated schedules, or planned future trajectories.  All dynamic coordination is local: each agent acts on its own state and on the observed relative states of nearby aircraft within a bounded sensing neighborhood. Details of the observed quantities and the policy architecture used are described in Section~\ref{sec:method}. We recognize that this represents a very strongly decentralized environment, and one that may sacrifice a significant amount of efficiency relative to a more centralized or federated setting. However, we believe that this investigation helps us understand what levels of performance may be achieved even in such a highly decentralized setting. 

\subsection{Communication requirements}
The state-observation model assumes that each aircraft can obtain sufficiently timely and accurate relative states of neighbors within its sensing range. In practice, this requires either cooperative position sharing (e.g., ADS-B broadcasts) or onboard sensing, together with an update rate compatible with the control timestep used by the policy.  We do not model communication dropouts, sensing latency, or positional uncertainty explicitly in this work; characterizing policy robustness under degraded surveillance, and quantifying the minimum update rate and sensing range required for safe operation, are important aspects that future work will have to address prior to deployment. 

\section{Methodology}
\label{sec:method}
In this section, we describe the environment and agent dynamics, detail the corridor navigation procedure, and describe the MARL training framework.
\subsection{Preliminaries}
\label{ssec:prelim}
Our environment builds on the InforMARL framework \cite{nayak_scalable_2023}, where aircraft are modeled as interacting agents operating in a shared airspace with goals and constraints.
We formulate the multi-aircraft system as a decentralized partially observable Markov decision process (Dec-POMDP) defined by the tuple 
$\langle N, S, O, \mathcal{A}, \mathcal{G}, P, R, \gamma \rangle$.
Each agent observes a local neighborhood and selects actions based on both state observations and graph-structured interaction information.

Specifically,
\begin{itemize}
    \item $N$ denotes the number of aircraft agents.
    \item $s \in S$ represents the global environment state.
    \item $o^{(i)} \in O$ is the local observation available to agent $i$.
    \item $a^{(i)} \in \mathcal{A}$ is the control action for agent $i$.
    \item $g^{(i)} \in \mathcal{G}$ is the interaction graph encoding nearby entities relative to agent $i$.
    \item $P(s'|s, A)$ defines the system transition dynamics under joint action $A$.
    \item $R(s,A)$ is the shared reward signal.
    \item $\gamma$ is the discount factor.
\end{itemize}

The objective is to learn decentralized policies 
$\Pi = (\pi^{(1)}, \ldots, \pi^{(N)})$, where each agent selects actions according to
$\pi^{(i)}(a^{(i)} \mid o^{(i)}, g^{(i)})$ using only local observations and interaction information.

\subsubsection{Agent dynamics}
\label{ssec:dynamics}
Aircraft motion is modeled using a planar fixed-wing kinematic model. Each agent state is defined as 
$s^{(i)} = [x, y, \theta, v]$, representing position, heading, and speed. The state evolves according to:
\begin{equation}
\dot{x}=v \cos \theta,\quad 
\dot{y}=v \sin \theta,\quad 
\dot{\theta}= \omega,\quad 
\dot{v} = a.
\end{equation}
where $\omega$ and $a$ are the control inputs corresponding to angular velocity and longitudinal acceleration, respectively.
The action for each agent is $a^{(i)} = [\omega^{(i)}, a^{(i)}]$, bounded by
$\omega \in [-\omega_{\max}, \omega_{\max}]$ and 
$a \in [a_{\min}, a_{\max}]$, with speed constrained to 
$v \in [v_{\min}, v_{\max}]$. Here, $a_{\min}<0$ represents maximum deceleration,
$a_{\max}>0$ maximum acceleration, and $v_{\min}, v_{\max}, \omega_{\max} > 0$
define the allowable speed range and symmetric turn-rate limits. The environment is integrated using forward Euler steps with a control timestep of $\Delta t = 1$ s, at which each agent receives an observation and issues a new action.

An agent is marked `done' upon reaching its assigned goal, at which point it is removed from further interaction. Episodes terminate when all agents complete their routes.

\subsection{Corridor navigation phases}

To analyze structured traffic flow, a single corridor is represented as a rectangular lane defined by length $L$, width $w$, and orientation $\theta^{\text{corridor}}$. Agent behavior relative to a corridor is divided into three conceptual phases $\phi \in \{1,2,3\}$:

\begin{enumerate}
    \item \textbf{Pre-corridor}: Agents navigate toward corridor entry points while maintaining separation and resolving any merge conflicts.
    \item \textbf{In-corridor}: Agents travel along the corridor while preserving alignment and longitudinal spacing.
    \item \textbf{Post-corridor}: Agents leave the corridor and diverge toward downstream goals.
\end{enumerate}

\begin{figure}[ht]
    \centering
    \includegraphics[width=0.5\columnwidth]{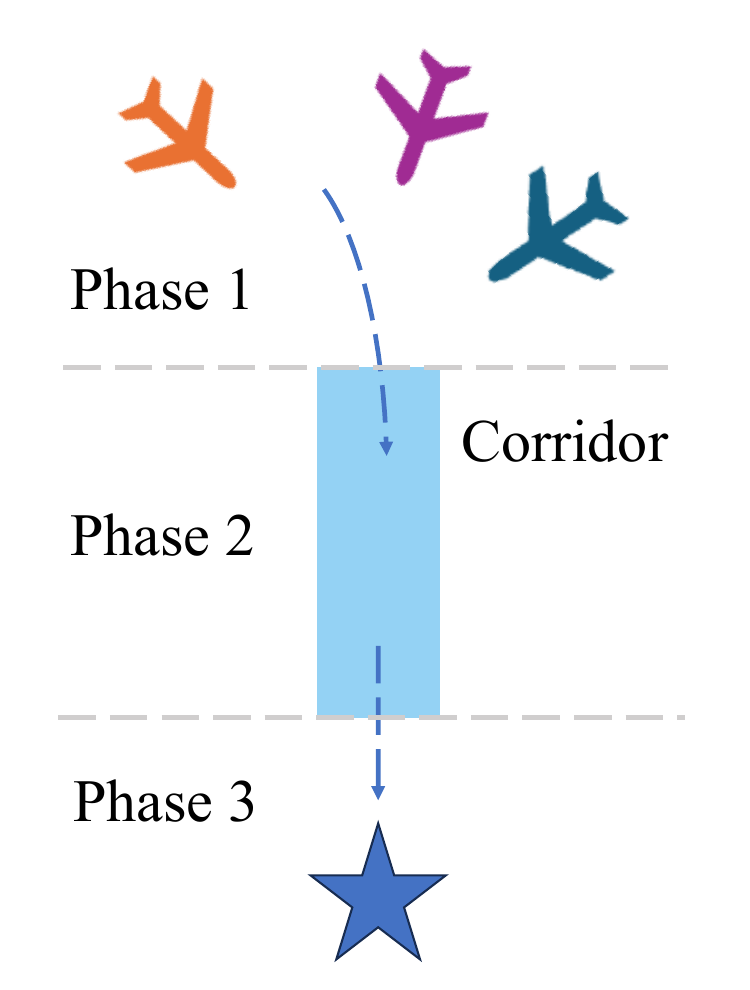}
    \caption{Single-corridor scenario (adapted from \cite{aloor_decentralized_2026}).}
    \label{fig:single_corridor_schematic}
\end{figure}

This phase-based description is used only for analysis and interpretation; the learned policy operates continuously across all phases. Figure \ref{fig:single_corridor_schematic} shows an example of a single corridor layout with the three phases.

\subsection{Multi-agent reinforcement learning framework}
Each aircraft is modeled as an autonomous agent that must traverse a designated corridor while maintaining separation from neighboring traffic and adhering to corridor constraints. Here, we describe the MARL methodology used to train agents for corridor coordination.

The objective is to learn a single decentralized policy that enables aircraft to:
\begin{itemize}
    \item maintain safe inter-aircraft separation,
    \item conform to corridor geometry and operational phases, and
    \item progress efficiently from entry to exit points.
\end{itemize}
Similar to \cite{aloor_decentralized_2026}, training is conducted in a single-corridor environment using a fixed number of agents per episode to learn stable coordination behaviors. Agent positions and headings are randomly initialized in every episode, and corridor lengths are sampled within the ranges in Table~\ref{tab:params}.

\subsubsection{Agent observations}
Each agent constructs a rotation-invariant local observation that remains consistent across corridor orientations and configurations. Expressing relative quantities in the agent’s heading frame removes dependence on global orientation, while the corridor-frame representation ensures consistency across differently aligned corridors. This design enables policies trained in a single-corridor setting to generalize to previously unseen geometries and layouts without the need to re-train the policy.
Each observation representation for any agent $o^{(i)}$ includes:
\begin{enumerate}
    \item ego state: heading $\theta$ relative to the corridor axis in the direction from entry to exit, and speed $v$
    \item goal position expressed in the ego frame $p^{\textrm{goal}}_i$
    \item relative position of the two nearest neighbors in the ego frame $p^{\textrm{n}_1}_i, p^{\textrm{n}_2}_i$
    \item normalized corridor-frame features:
        \begin{enumerate}
            \item longitudinal progress along the corridor $s$
            \item lateral deviation from centerline $d^{\textrm{center}}/w$
            \item normalized distances to entrance ($d^{\textrm{entrance}}/L$) and exit gates ($d^{\textrm{exit}}/L$)
            \item nearest-neighbor distance and closing-in rate
            \item phase of corridor traversal indicator $\phi \in \{1,2,3\}$
        \end{enumerate}
\end{enumerate}
The final ego observation vector $o^{(i)}$ received by an agent can be represented as 
\begin{equation}
\begin{aligned}
o^{(i)} = \big[
&\theta_i,\; v_i,\; p^{\mathrm{goal}}_i,\; p^{\mathrm{n}_1}_i,\; p^{\mathrm{n}_2}_i,\; s, 
\tfrac{d^{\mathrm{center}}_i}{w},\;
\tfrac{d^{\mathrm{entrance}}_i}{L},\;
\tfrac{d^{\mathrm{exit}}_i}{L},\\
&d^{\mathrm{n}_1}_i,\;
\dot{d}^{\mathrm{n}_1}_i,\;
\phi_i
\big].
\end{aligned}
\end{equation}

\subsubsection{Reward function design}
The reward structure is designed to produce stable and scalable behaviors across corridor geometries and traffic densities. The reward combines formation maintenance, corridor adherence, and progress incentives. Key components include:
\begin{enumerate}
    \item \textit{Separation maintenance:} Agents are penalized when inter-agent spacing falls below the desired separation threshold, stronger penalties are applied when aircraft are both within the separation threshold and approaching one another (positive closing rate), indicating increasing conflict risk.
    \item \textit{Corridor adherence and phase progression:} Agents receive shaping rewards encouraging entry into the corridor, alignment with the corridor axis, and forward progression toward the exit. Structured bonuses are provided when agents correctly transition between the pre-corridor, in-corridor, and post-corridor phases, while skipped or reversed transitions incur penalties.
    \item \textit{Goal completion:} A terminal reward is provided upon reaching the goal, only after traversing through the corridor.
\end{enumerate}
To improve training stability and encourage robust corridor conformance, reward terms and collision penalties are introduced progressively using a curriculum learning strategy. Early training emphasizes corridor adherence and phase progression to establish stable formation and navigation behaviors. As policies mature, separation penalties are gradually increased, and collision penalties are raised to a higher magnitude to enforce separation maintenance. Policies were trained using the parameters in Table \ref{tab:params}. Single-corridor performance under similar evaluation densities was characterized in our prior work \cite{aloor_decentralized_2026}.

\begin{table}[h]
\centering
\caption{Parameter values during training}
\begin{tabular}{l l}
\hline
\textbf{Parameter} & \textbf{Value} \\
\hline
Minimum groundspeed, $v_{\text{min}}$ & 60 knots (111 km/h) \\
Maximum groundspeed, $v_{\text{max}}$ & 175 knots  (324 km/h) \\
Minimum acceleration, $a_{\text{min}}$ & $-1$ m/s$^2$ \\
Maximum acceleration, $a_{\text{max}}$ & 2 m/s$^2$ \\
Maximum angular velocity, $\omega_{\max}$ & 0.075 rad/s (4.27 deg/s)\\
Length of air corridor & 1.2 km to 2 km \\
Width of air corridor & 0.2 km \\
Minimum inter-aircraft separation & 0.15 km \\
Goal threshold distance & 0.35 km \\
Episode length & 150 s \\
Number of agents in training & 5 \\
\hline
\end{tabular}
\label{tab:params}
\end{table}

\section{Evaluation setup}
\label{sec:experiment}
This section evaluates the performance of the learned decentralized coordination policy when deployed in corridor networks more complicated than the training configuration. The objective is to assess scalability, generalization across corridor geometries, and the resulting traffic flow characteristics under increasing operational complexity and demand. Once the model is trained in the single-corridor environment, it is frozen and evaluated without further learning. The resulting policy operates solely on local observations and learned interaction representations, enabling decentralized execution without centralized traffic management or global state information.

All experiments are designed as an online navigation task in which agents are not provided with the full corridor sequence in advance. Instead, each corridor is revealed incrementally as the agent exits the previous one. At each transition, the observation space is updated to include only local geometric information about the next corridor, specifically its entrance location, orientation, and relative distance with respect to the agent. As a result, agents must make decisions using locally available information rather than a precomputed global route. The task, therefore, evaluates the learned policy’s ability to adapt to newly encountered corridor configurations and maintain coordination under sequential, partially observable conditions.

Traffic demand is varied by adjusting the number of aircraft operating simultaneously in the scenario. Experiments are conducted with 10, 20, 30, and 40 aircraft, representing increasing traffic density and operational complexity. Aircraft enter the network over time according to a fixed metering rate, producing sustained traffic flows and enabling the evaluation of throughput and congestion propagation. 
These scenarios enable evaluation of both localized coordination, such as those in merges, and system-level traffic flow behavior across networked corridors. Evaluation parameters can be found in Table \ref{tab:eval_params}.

\begin{table}[h]
\centering
\caption{Parameters used for evaluation.}
\begin{tabular}{p{4cm} p{4cm}}
\hline
\textbf{Parameter} & \textbf{Value} \\
\hline
Episode length & 200--500 s depending on network size and number of aircraft \\
Number of aircraft & 10, 20, 30, and 40 \\
Length of air corridor & 1.2 km to 4 km \\
Number of experiment trials & 50 (for 10-30 agents), 10 (for 40 agents) \\
Heterogeneous performance case & A subset of agents is randomly selected and limited to a reduced maximum speed of 140 knots, while remaining agents retain $v_{\text{max}}=175$ knots \\
\hline
\end{tabular}
\label{tab:eval_params}
\end{table}



\subsection{Performance metrics}
\label{ssec:metrics}
To evaluate decentralized coordination at the system level, we consider a set of operational and flow-oriented performance indicators. These metrics quantify navigation reliability, traffic efficiency, exposure to potential collision, and network throughput under increasing demand and corridor complexity. All metrics are averaged across aircraft and trials for each scenario.
\subsubsection{Conformance to corridor boundaries ($C\%$)}
This metric measures how well aircraft stay within the specified boundary walls of the corridors. Conformance is calculated as an average percentage of time an aircraft remains within the boundary while traversing through the corridor, while in phase 2.
\subsubsection{Completion rate ($S\%$)}
This metric tracks the percentage of aircraft that successfully navigate the sequence of corridors assigned to them within the length of the episode. The higher the completion rate, the better the task performance. 
\subsubsection{Average speed (knots)}
We calculate the average speed by measuring the total distance each agent travels along the corridor network and dividing it by the time taken.
\subsubsection{Need for tactical intervention for deconfliction ($I\%$)}
At any timestep when an aircraft is too close to another aircraft and violates the minimum separation distance, a tactical intervention for deconfliction is needed and is flagged. We measure the need for tactical intervention ($I\%$) as the fraction of time an agent requires tactical intervention over the total time it takes to traverse a corridor. 

\section{Results and Discussion}
\label{sec:results}
Both the physical environment and temporal horizon are varied across scenarios to evaluate scalability. The spatial extent of the environment is scaled with the number of corridors and aircraft to maintain realistic corridor densities while enabling evaluation under larger operational areas. Similarly, the episode length is adjusted based on network size and traffic demand to allow sufficient time for aircraft to traverse corridors and for traffic flow patterns to stabilize. 






\subsection{Homogeneous traffic performance}
\label{ssec:homogen}
We first analyze system behavior under homogeneous operations, where all aircraft share identical performance envelopes, i.e., the same $v_{\text{max}}=175$ knots. We evaluate decentralized coordination across a set of corridor network topologies representing progressively increasing operational complexity.
\subsubsection{Simple network scenarios}
\label{ssec:scenarios}
\begin{figure}[ht!]
    \centering
    \includegraphics[width=0.7\linewidth,keepaspectratio]{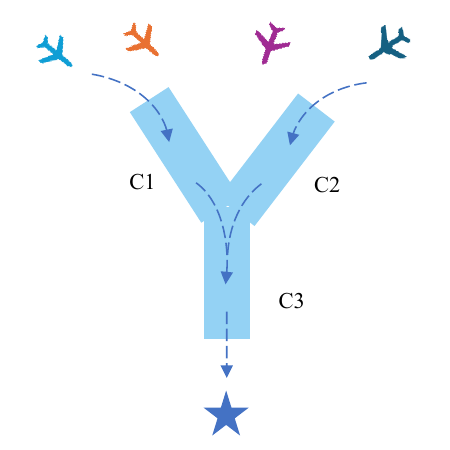}
    \caption{Merge corridor geometry.}
    \label{fig:merge}
\end{figure}

\noindent\paragraph{Merge} Two upstream corridors merge into a single downstream corridor. This scenario captures the capacity constraints and interaction dynamics at merge points and consists of 3 corridors. (See Fig. \ref{fig:merge}).


\begin{figure}[ht!]
    \centering
    \includegraphics[width=0.6\linewidth,keepaspectratio]{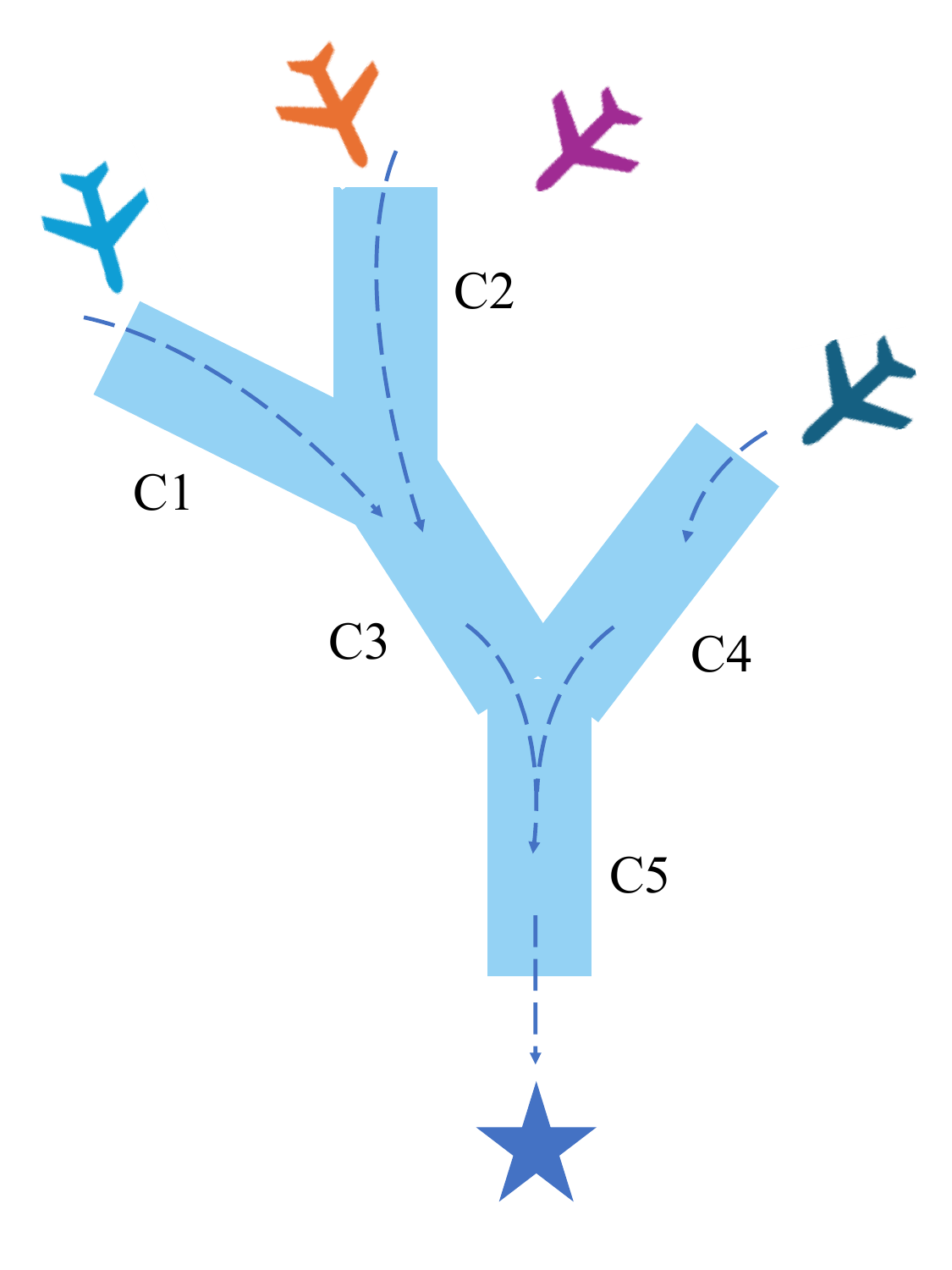}
    \caption{Double-merge corridor geometry.}
    \label{fig:doublemerge}
\end{figure}

\begin{table}[ht!]
\caption{Performance metrics for the single-merge corridor scenario. Metrics defined in Sec.~\ref{ssec:metrics}.}
\centering
\begin{tabular}{|c||c|c|c|c|}
\hline
\multirow{2}{*}{\textbf{\#}} & \multirow{2}{*}{\shortstack{Conformance\\$C\%$}} & 
\multirow{2}{*}{\shortstack{Completion rate\\$S\%$}} &
\multirow{2}{*}{\shortstack{Avg. speed\\ (knots)}} &
\multirow{2}{*}{\shortstack{Tactical\\intervention $I\%$}} \\
 & & & & \\
\hline \hline
10 & 99\% & 99\% & 173.4 & 1.25\%\\
20 & 99\% & 99\% & 173.3 &  1.64\%\\
30 & 99\% & 99\% & 173.1 &  1.83\%\\
40 & 99\% & 90\% & 174.5 &  1.48\%\\
\hline
\end{tabular}
\label{tab:merge}
\end{table}
\noindent\paragraph{Double merge} Two sequential merging points create compounded interaction effects and increased local traffic density. This scenario includes 5 corridor segments. (See Fig. \ref{fig:doublemerge}).

\begin{figure}[h!]
    \centering
    \includegraphics[width=0.33\linewidth,keepaspectratio]{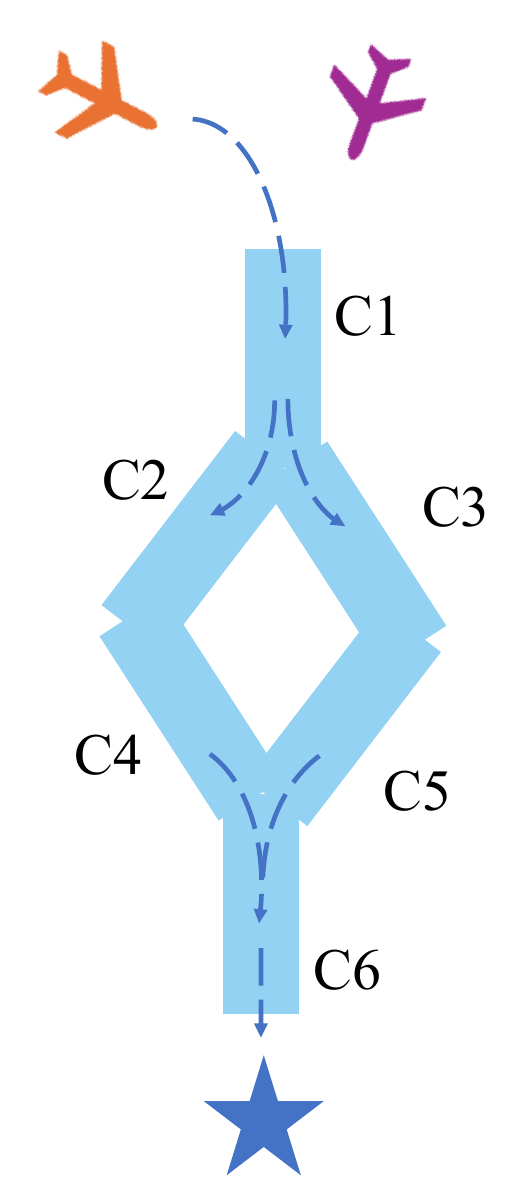}
    \caption{Split--merge corridor geometry.}
    \label{fig:splitmerge}
\end{figure}

\begin{table}[ht!]
\caption{Performance metrics for the double-merge corridor scenario.}
\centering
\begin{tabular}{|c||c|c|c|c|}
\hline
\multirow{2}{*}{\textbf{\#}} & \multirow{2}{*}{\shortstack{Conformance\\$C\%$}} & 
\multirow{2}{*}{\shortstack{Completion rate\\$S\%$}} &
\multirow{2}{*}{\shortstack{Avg. speed\\ (knots)}} &
\multirow{2}{*}{\shortstack{Tactical\\intervention $I\%$}} \\
 & & & & \\
\hline \hline
10 & 99\% & 96\% & 171.9 & 4.76\%\\
20 & 99\% & 92\% & 172.2 & 4.63\%\\
30 & 98\% & 89\% & 172.3 & 5.16\%\\
40 & 98\% & 88\% & 168.2 & 5.78\%\\
\hline
\end{tabular}
\label{tab:double_merge}
\end{table}
\noindent\paragraph{Split and merge}
Aircraft first diverge into parallel corridors and subsequently merge into a single downstream flow, introducing route-choice and synchronization challenges. This scenario includes 6 corridors. (See Fig. \ref{fig:splitmerge}).
\begin{table}[h!]
\caption{Performance metrics for the split-and-merge corridor scenario.}
\centering
\begin{tabular}{|c||c|c|c|c|}
\hline
\multirow{2}{*}{\textbf{\#}} & \multirow{2}{*}{\shortstack{Conformance\\$C\%$}} & 
\multirow{2}{*}{\shortstack{Completion rate\\$S\%$}} &
\multirow{2}{*}{\shortstack{Avg. speed\\ (knots)}} &
\multirow{2}{*}{\shortstack{Tactical\\intervention $I\%$}} \\
 & & & & \\
\hline \hline
10 & 99\% & 99\% & 173.7 & 0.61\%\\
20 & 99\% & 98\% & 171.2 & 0.57\%\\
30 & 99\% & 98\% & 173.9 & 0.52\%\\
40 & 99\% & 98\% & 174.6 & 0.97\%\\
\hline
\end{tabular}
\label{tab:split_merge}
\end{table}
Across all scenarios, corridor conformance remains consistently high, exceeding 96\% even at the highest traffic levels. This indicates that the learned coordination policies maintain stable alignment with corridor structures despite increased interaction density and geometric complexity.

Completion rates are also very high, almost 100\% in low- and medium-density cases, but decrease under higher demand, particularly in the double-merge scenario.
This reduction is primarily driven by localized congestion and increased interaction frequency near merging points, resulting in some aircraft deviating significantly from their route in order to maintain separation. Consequently, especially for aircraft that are later in the flow, they may not be able to reach their exit before the episode ends. 

Average speeds remain relatively stable across traffic levels, with only modest reductions in the most interaction-heavy configuration of double-merge. This suggests that agents prioritize maintaining forward progress and flow efficiency even under increased coordination demands, rather than adopting conservative slowdown behaviors. 

The need for tactical intervention increases with traffic density and corridor complexity, most noticeably in double-merge configurations where interaction zones are spatially concentrated. However, intervention rates remain low overall, indicating that separation conflicts are largely managed through learned strategic coordination rather than reactive corrections. The lowest need for tactical intervention rates is in the split-merge scenario, as the merge point only encounters half the density it sees in the other scenarios.
\begin{figure}[ht!]
    \centering
    \includegraphics[width=\linewidth]{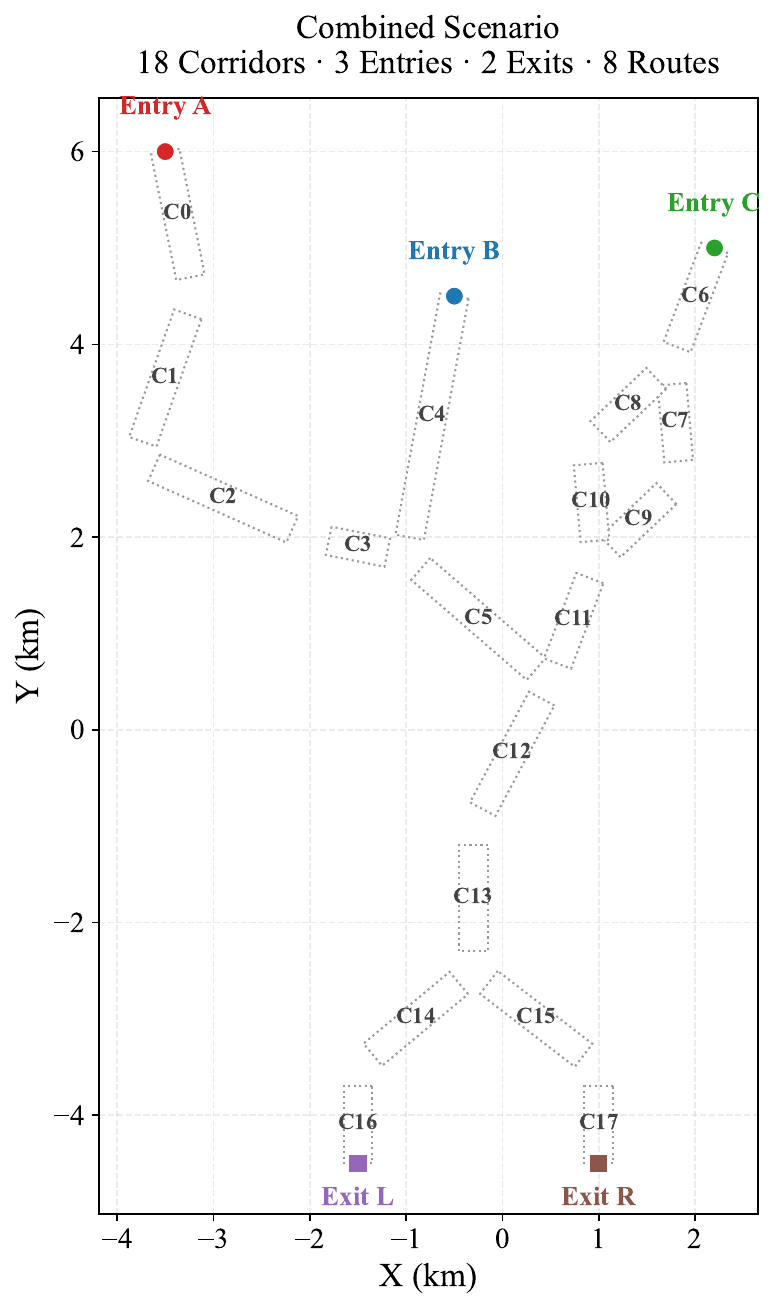}
    \caption{Combined corridor scenario built using merges, splits, and multiple sequential chains of corridors totaling 18}
    \label{fig:combined18corridor}
\end{figure}

\subsubsection{Combined corridor network with homogeneous traffic}
We now test our trained policy on the large-scale corridor network shown in Figure \ref{fig:SFBaycorriday}, composed of multiple merges, splits, and sequential segments, totaling up to 18 interconnected corridors (see Figure \ref{fig:combined18corridor} for a more detailed description). This scenario reflects realistic high-density AAM operations with distributed coordination requirements. Despite this increased structural and interaction complexity, the decentralized policy maintains strong performance across all traffic levels. Corridor conformance remains high, decreasing only slightly from 98\% to 96\% as traffic increases, indicating that agents preserve stable route-following behavior even when navigating long multi-segment trajectories. Completion rates remain above 97\% across all demand levels, demonstrating that agents are able to traverse extended corridor sequences without prior knowledge of the full network and while relying only on local corridor information. Average speeds remain consistent across traffic densities, suggesting that flow efficiency is preserved even as interaction complexity grows and tactical intervention rates remain moderate. 
\begin{figure*}[t!]
\centering
\begin{subfigure}{0.28\linewidth}
    \centering
    \includegraphics[width=\linewidth]{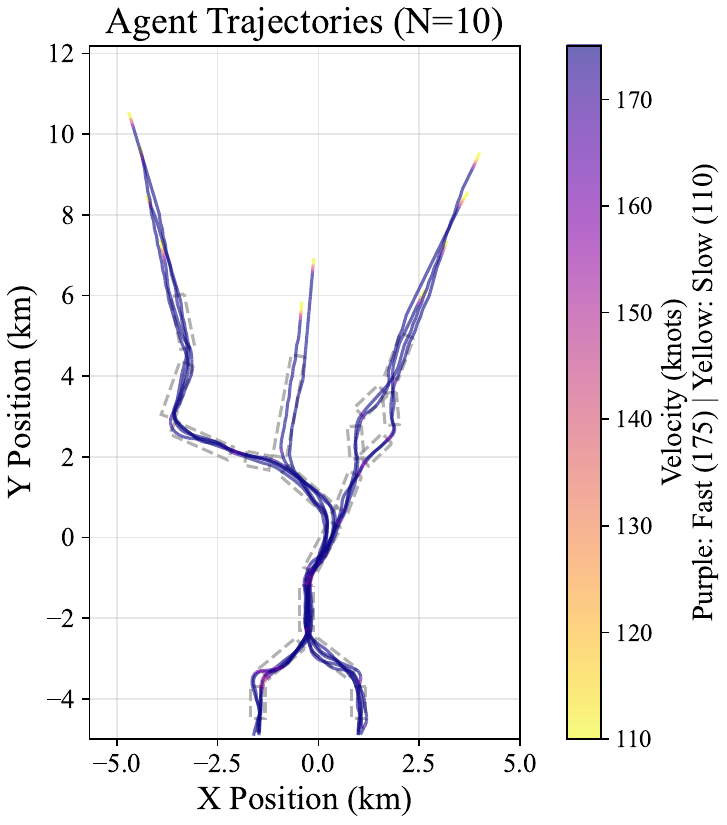}
    \caption{Homogeneous agent trajectories}
    \label{fig:trajhomogeneousspeeds}
\end{subfigure}
\begin{subfigure}{0.21\linewidth}
    \centering
    \includegraphics[width=\linewidth]{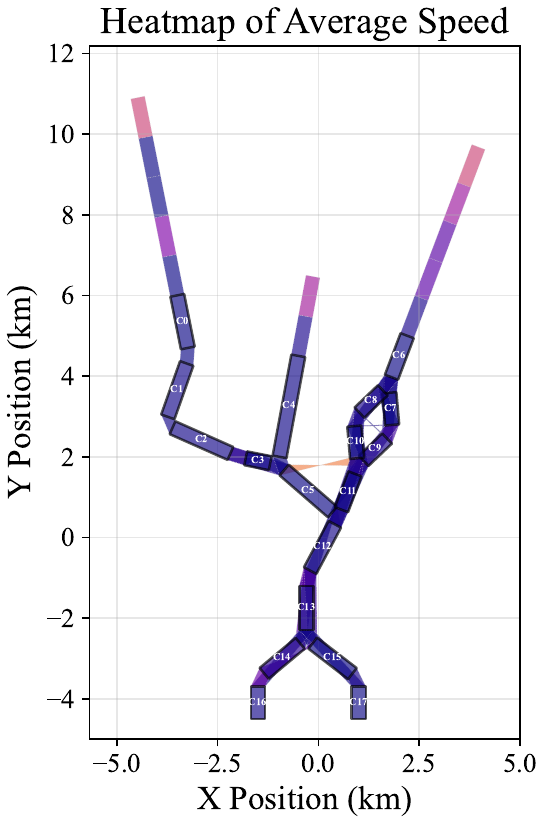}
    \caption{Homogeneous speed heatmap}
    \label{fig:maphomogeneousspeeds}
\end{subfigure}
\begin{subfigure}{0.46\linewidth}
    \centering
    \includegraphics[width=\linewidth]{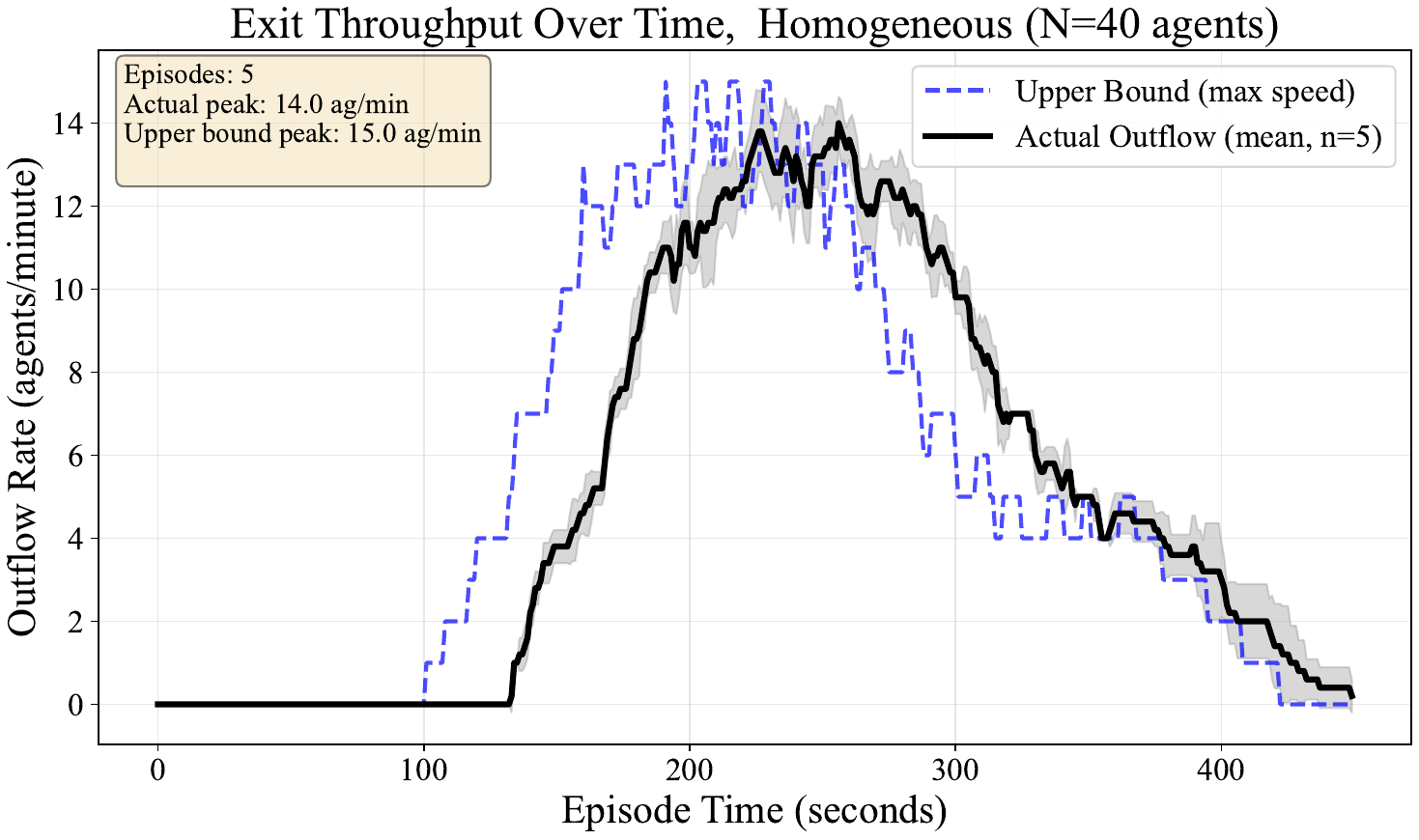}
    \caption{Exit throughput of the combined corridor scenario}
    \label{fig:outflowrate_uniform}
\end{subfigure}
\caption{(a, b) AAM corridor navigation performance in the combined corridor scenario of 18 total corridors. The plots show each agent's trajectory colored by instantaneous velocity, along with a spatial heatmap of average agent velocities across corridor segments and gaps. Approach zones before entry corridors (C0, C4, C6) show initial speeds before corridor entry. Homogeneous agents with $v_{\text{max}} =175$ knots, and $N=10$ agents per scenario. Purple represents fast (175 knots or 324 km/hr), and yellow represents slow (110 knots or 204 km/hr). (c) Exit throughput of the combined corridor scenario (18 corridors) over episode time when all agents have the same speed limits, comparing actual policy performance (solid black) against theoretical upper bound on throughput (dashed blue). Upper bound computed assuming all agents travel at maximum speed (175 knots) along their assigned routes. Sliding window: 60 seconds. }
\label{fig:traj_heatmap}
\end{figure*}

\begin{table}[h!]
\caption{Performance metrics for the combined corridor network scenario.}
\centering
\begin{tabular}{|c||c|c|c|c|}
\hline
\multirow{2}{*}{\textbf{\#}} &
\multirow{2}{*}{\shortstack{Conformance\\$C\%$}} &
\multirow{2}{*}{\shortstack{Completion rate\\$S\%$}} &
\multirow{2}{*}{\shortstack{Avg. speed\\ (knots)}} &
\multirow{2}{*}{\shortstack{Tactical\\intervention $I\%$}} \\
 & & & & \\
\hline \hline
10 & 98\% & 99\% & 171.4 & 3.7\%  \\
20 & 98\% & 99\% & 171.9 & 4.2\%  \\
30 & 97\% & 98\% & 172.8 & 3.6\%  \\
40 & 96\% & 97\% & 173.2 & 3.3\%  \\
\hline
\end{tabular}
\label{tab:combined}
\end{table}
\begin{table}[htbp]
\centering
\caption{Statistics by route for the combined corridor scenario with 40 agents. Routes are labeled by entry point (A, B, C), exits (L, R); and intermediate corridors at splits for disambiguation.} \label{table:route_speeds}
\begin{tabular}{|l||l|c|c|c|}
\hline 
Rte & Entry$\rightarrow$Exit & Avg. shortest & Avg. actual & Avg. speed \\
 & & distance (km) & distance (km) & (knots) \\
\hline \hline
R1 & A$\rightarrow$L & 13.8 & 14.6 & 174.2 \\
R2 & A$\rightarrow$R & 13.8 & 14.4 & 170.0 \\
R3 & B$\rightarrow$L & 10.3 & 11.3 & 163.0 \\
R4 & B$\rightarrow$R & 10.3 & 11.1 & 167.0 \\
R5 & C$\rightarrow$C8$\rightarrow$ L & 10.7 & 11.8 & 155.3 \\
R6 & C$\rightarrow$C8$\rightarrow$R & 10.7 & 11.7 & 162.4 \\
R7 & C$\rightarrow$C7$\rightarrow$ L& 10.7 & 11.6 & 163.5 \\
R8 & C$\rightarrow$C7$\rightarrow$ R & 10.7 & 11.8 & 169.2 \\
\hline
\multicolumn{4}{l}{\footnotesize Max speed: 175.0 knots} \\
\end{tabular}
\end{table}

Route-level statistics are shown in Table \ref{table:route_speeds}, showing how flow is distributed over the various corridors. Even in this congested and complex corridor scenario, average speed only deviates from the maximum speed by $5.4\%$, and the actual distance traveled is within $8.02\%$ of the shortest possible routes (without considering congestion). Similarly, the average time deviation is within $13.5\%$ of the fastest possible travel times, in the absence of congestion. These bounds illustrate the efficiency of the learned traffic management policies. 

The homogeneous setting also provides a baseline estimate of the achievable network capacity under idealized operating conditions. To quantify this, we evaluate the exit throughput of the combined corridor network in Figure \ref{fig:combined18corridor} under the 40-agent case. Figure \ref{fig:outflowrate_uniform} shows the time-varying outflow rate compared against a theoretical upper bound on the throughput computed by assuming all aircraft travel at their maximum allowable speed (175 knots) along conflict-free trajectories. The learned policy closely tracks the theoretical limit, reaching a peak throughput of approximately 14 aircraft per minute compared to the upper bound value of 15 aircraft per minute. The small gap between achieved and theoretical throughput arises primarily from localized coordination delays at merge junctions and transient speed adjustments needed to maintain separation and navigate turns.

\begin{figure*}[ht!]
\centering
\begin{subfigure}{0.28\linewidth}
    \centering
    \includegraphics[width=\linewidth]{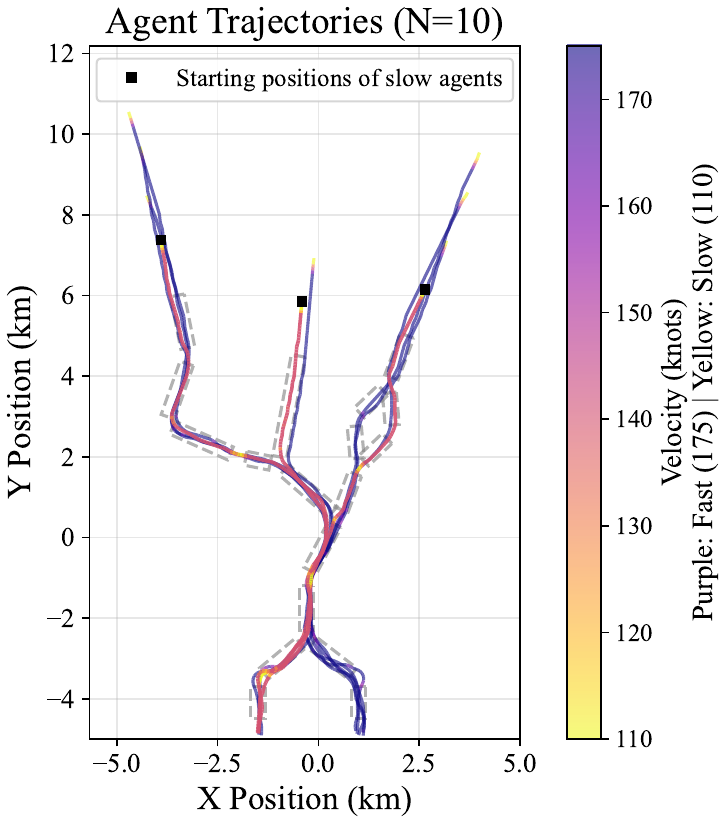}
    \caption{Heterogeneous agent trajectories}
    \label{fig:trajhetgeneousspeeds}
\end{subfigure}
\hfill
\begin{subfigure}{0.21\linewidth}
    \centering
    \includegraphics[width=\linewidth]{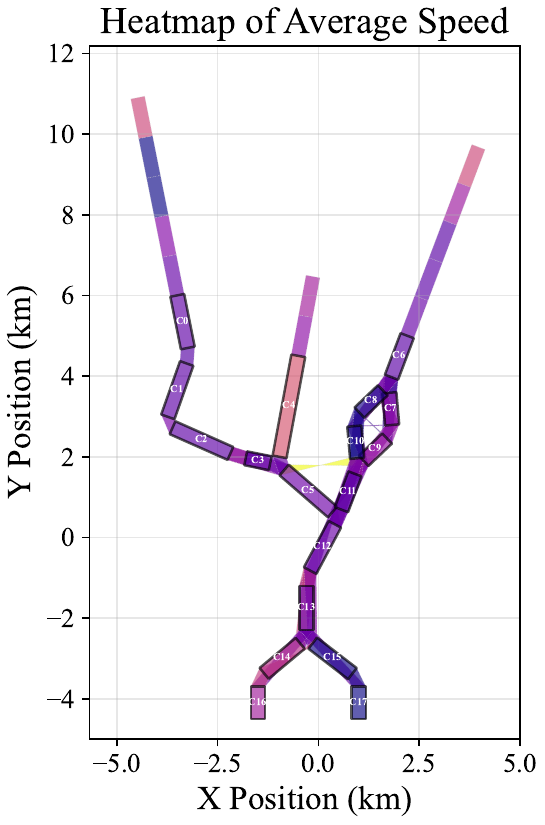}
    \caption{Het. speed heatmap}
    \label{fig:maphetgeneousspeeds}
\end{subfigure}
\begin{subfigure}{0.46\linewidth}
    \centering
    \includegraphics[width=\linewidth]{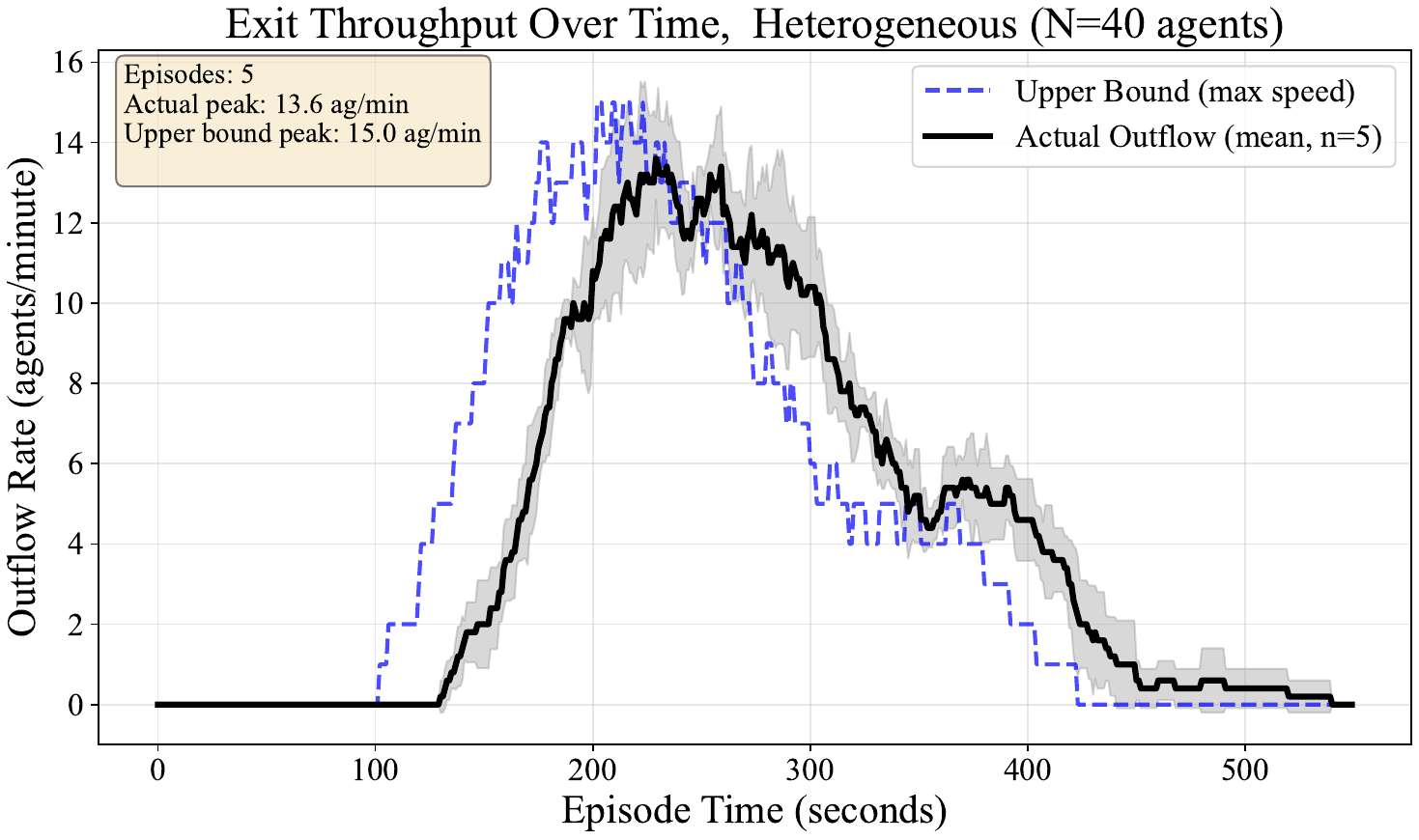}
    \caption{Exit throughput of the combined corridor scenario (heterogeneous)}
    \label{fig:outflowrate_het}
\end{subfigure}
\caption{(a, b) AAM corridor navigation performance in the combined corridor scenario of 18 total corridors with heterogeneous speed agents (three agents at $v_{\text{max}}$ that is 20\% slower, i.e., 140 knots or 259 km/hr). The plots show each agent's trajectory colored by instantaneous velocity, along with a spatial heatmap of average agent velocities across corridor segments and gaps. Approach zones before entry corridors (C0, C4, C6) show initial speeds before corridor entry. N=10 agents per scenario. Purple represents fast (175 knots or 324 km/hr), and yellow represents slow (110 knots or 204 km/hr). Black squares mark the starting positions of slower agents. (c) Exit throughput of the combined corridor scenario (18 corridors) over episode time. Actual policy performance (solid black) is compared against theoretical upper bound on the throughput (dashed blue), computed assuming all agents travel at 175 knots. Sliding window: 60 seconds. }
\label{fig:traj_heatmap_heterogen}
\end{figure*}

Trajectory visualizations provide additional insight into how agents negotiate dense traffic. The instantaneous speeds of each agent in a 10-agent episode along the entire trajectory and the average speed across various corridor segments are shown in Fig. \ref{fig:trajhomogeneousspeeds} and Fig. \ref{fig:maphomogeneousspeeds}. Aircraft maintain high cruise speeds within corridors with a small reduction in speed while making turns along inter-corridor gaps and near merge junctions where interaction density increases. In  particular, we see that aircraft try to slow down or stretch their paths in between corridors to maintain separation, especially just before entering the last corridor (which tends to be among the most congested). Such localized slowdowns and path stretches reflect coordination maneuvers rather than global congestion, and aircraft reaccelerate once they are clear of these regions of increased local interactions.

\subsection{Heterogeneous traffic performance}
\label{ssec:heterogen}

To reflect realistic AAM operations involving aircraft with varying performance envelopes, we conducted experiments with heterogeneous vehicle capabilities. Multiple maximum ground speed ($v_{\max}$) classes are introduced by giving different speed constraints to agents. This heterogeneity results in mixed-speed traffic flows, requiring decentralized coordination mechanisms to accommodate differences in maneuverability and traversal time while maintaining separation and corridor conformance.
To evaluate robustness under more realistic operations, we introduce performance heterogeneity by assigning a subset of aircraft a reduced maximum speed of 140 knots while the remaining aircraft retain a maximum speed of 175 knots. This creates mixed-performance traffic streams similar to real AAM operations, where vehicle capabilities and mission constraints vary.
We recompute the exit throughput for the combined corridor network under this heterogeneous configuration. Compared with the homogeneous case, throughput decreases slightly due to the presence of these slower aircraft acting as moving constraints within high-density segments. Fig. \ref{fig:traj_heatmap_heterogen} shows the corridor navigation performance for a scenario with 10 heterogeneous agents. 

Trajectory and speed visualizations reveal emergent coordination behaviors in the heterogeneous case. Faster agents adjust their speeds when approaching slower aircraft within narrow corridor segments, but opportunistically overtake in inter-corridor gaps and lateral buffer regions where interaction constraints are reduced. This behavior is particularly evident near sequential merge locations, where traffic density is highest, and multiple routing options exist. As a result, localized speed reductions occur upstream of slow agents, while downstream flow recovers as faster aircraft re-accelerate after passing. We find the emergence of this behavior, noticeable in the bottom-right of Fig. \ref{fig:trajhetgeneousspeeds}, quite interesting: aircraft learn to wait until they are outside a corridor to extend their paths while overtaking slower aircraft, thereby avoiding the violation of corridor boundaries. As a result, the presence of slower aircraft does not lead to persistent congestion or breakdown of coordination. Instead, the system exhibits adaptive speed harmonization and opportunistic passing, indicating that the learned policy captures interaction-aware navigation rather than rigid speed-following.

\section{Limitations and Directions for Future Work}
\label{sec:limit}
While this study demonstrates the effectiveness of decentralized MARL policies for coordination in structured corridor networks, some limitations remain that motivate future research.

\paragraph{Tactical safety layer}
While our prior work considered safety filters along with MARL in a layered manner \cite{AloorJ-RSS-25}, in this work, separation is enforced through reward shaping and collision penalties rather than formal guarantees. Extending the framework with hard safety constraints, such as runtime safety filters \cite{AloorJ-RSS-25, ahmad2026hmarl, low2026timetoreach}, would improve resilience to disturbances, sensing uncertainty, and unexpected failure modes.

\paragraph{Off-nominal conditions}
Our numerical experiments were all conducted under nominal operating conditions, in which all aircraft remained controllable and follow their assigned routes. Off-nominal events---such as vehicle failures, communication loss or unplanned route deviations due to weather---were not modeled here. Ablation studied that evaluate whether the learned coordination behaviors are robust to communication dropouts and whether they degrade gracefully under such contingencies is an important open question, and will need to be addressed prior to operational deployment.

\paragraph{Environmental uncertainty}
The environment considered in this paper did not include wind or other environmental disturbances. Furthermore, all agents were assumed to share the same idealized observation model. In realistic operations, merging aircraft may have different and evolving information about local wind conditions and may experience disturbances that affect the separation minima. Extending the framework to handle environmental uncertainties, and allowing aircraft to coordinate effectively under heterogeneous or partial information about ambient conditions, is an interesting topic for future work.

\paragraph{Fixed corridor structure and planar dynamics}
The present work assumed fixed corridor geometries and routes. Future research will investigate dynamic, on-demand corridor assignment and adaptive routing strategies that respond to changing traffic demand and operational constraints.  Furthermore, the experiments conducted to date focused on planar motion and simplified vehicle dynamics. Incorporating fully three-dimensional operations, along with more realistic vehicle performance and environmental effects, will be important for operational relevance. Other interesting directions include the integration with terminal-area airspaces and a hybrid centralized-decentralized architecture. 

\paragraph{Implications for deployment of AAM systems}
Deployment of decentralized learned coordination will require complementary tactical safety layers and communication infrastructure sufficient to support local state observation at the required update rates, and clearly defined operational procedures for contingencies and degraded conditions. We see characterizing these requirements quantitatively and co-designing the learning framework with them in mind as a promising direction for future work, and one that begins to answer the question of what the needed equipage, infrastructure, and procedural standards would be for decentralized AAM operations.

\section{Conclusions}
\label{sec:conc}
This paper presented a decentralized MARL framework for autonomous aircraft coordination in structured AAM corridor networks across varying traffic densities and corridor topologies. Using a rotation-invariant policy trained to navigate a single corridor, we demonstrated generalization to multi-corridor configurations, maintaining high conformance and stable traffic flow as interaction complexity increased. Evaluations included merges, splits, and combined-graph scenarios of 18 corridors with up to 40 aircraft at a time, including heterogeneous fleets. Our results showed that decentralized autonomy can function as an effective strategic traffic flow management mechanism, enabling coordination across structured AAM corridor networks.

\section*{Acknowledgment}
The authors would like to thank the MIT SuperCloud \cite{supercloud} and the Lincoln Laboratory Supercomputing Center for providing high-performance computing resources that have contributed to the research results reported within this paper.

\bibliographystyle{ieeetr}
\bibliography{references}
\vspace{12pt}

\flushend

\end{document}